\documentclass[11pt,twoside]{article}
\usepackage{asp2004}
\usepackage{psfig}
\usepackage{epsf}
\usepackage{graphics}
\usepackage{lscape}
\def\edcomment#1{\iffalse\marginpar{\raggedright\sl#1\/}\else\relax\fi}
\reversemarginpar

\begin{document}
\title{The Ly$\beta$ and O\,VI Forest in the Local Universe}
\author{Charles W. Danforth \& J. Michael Shull}
\affil{University of Colorado, 389-UCB, Boulder, CO 80309, USA}

\begin{abstract}
Intergalactic absorbers along lines of sight to distant quasars are a powerful diagnostic for the evolution and content of the IGM.  In this study, we search known low-$z$ Ly$\alpha$ absorption systems for equivalent absorption in higher Lyman lines as well as the important metal ions O\,VI and C\,III.  O\,VI absorption is detected roughly half the time and shows a smaller range of column density than H\,I.  This implies a multi-phase IGM with warm neutral and hot, ionized components.  We use our O\,VI detection statistics to determine that the fraction of intergalactic mass at $10^{5-6}$ K is at least 5\% of the total baryonic mass in the local universe.
\end{abstract}
\thispagestyle{plain}

\section{Introduction}
Measurements of the low-redshift intergalactic medium (IGM) are critical to our understanding of the large scale structure and ionization state of the local universe as well as the interaction of galaxies with their environment.  FUSE is ideally suited to measure intergalactic H\,I absorption up to $z\rm_{abs}<0.3$, O\,VI to $z\rm_{abs}<0.15$, and C\,III to $z\rm_{abs}<0.21$.

We started with published lists of intervening Ly$\alpha$ absorption systems toward low-$z$ AGNs, obtained from GHRS and STIS surveys by \citet{Penton1,Penton4}.  Other sight lines were covered by literature sources or measured at the University of Colorado from STIS/E140M spectra.  We disregarded any weak Ly$\alpha$ absorbers (W$_\lambda<80$ m\AA, log\,N(H\,I)$\la13.2$) and searched the FUSE data for higher-order Lyman lines, O\,VI, and C\,III counterparts. We measured all available Lyman series lines to determine N and $b$ values for H\,I via curves of growth.  Metal-ion columns were determined via profile fits; we assume that any saturation in these lines will be mild, and that profile fits accurately determine the column density. 

We measured 167 absorbers with W$_{\rm\lambda}$(Ly$\alpha$)$>80$ m\AA\ toward 26 AGN sight lines.  Of these absorbers, 126 were at $z\rm_{abs}<0.15$ with 52 O\,VI 3$\sigma$ detections in one or both lines of the doublet and 63 well-defined upper limits.  The O\,VI lines for the remaining 11 absorbers fall on top of airglow lines, strong ISM lines, or are in some other way inaccessible.  For absorbers where both lines of the O\,VI doublet were measured, we calculated a weighted mean for N(O\,VI).  We find that the differential number  $\cal{N}$ of O\,VI absorbers detected with column density N$_{OVI}$ is consistent with a power law with index $-2.4$: $\cal{N}\rm (N_{OVI})\propto N_{OVI}^{-2.4}$.  This is much steeper than the equivalent power-law relationship for H\,I:  $N_{HI}^{-1.65\pm0.07}$ \citep{Penton4}.

\begin{figure}[t]
\plotone{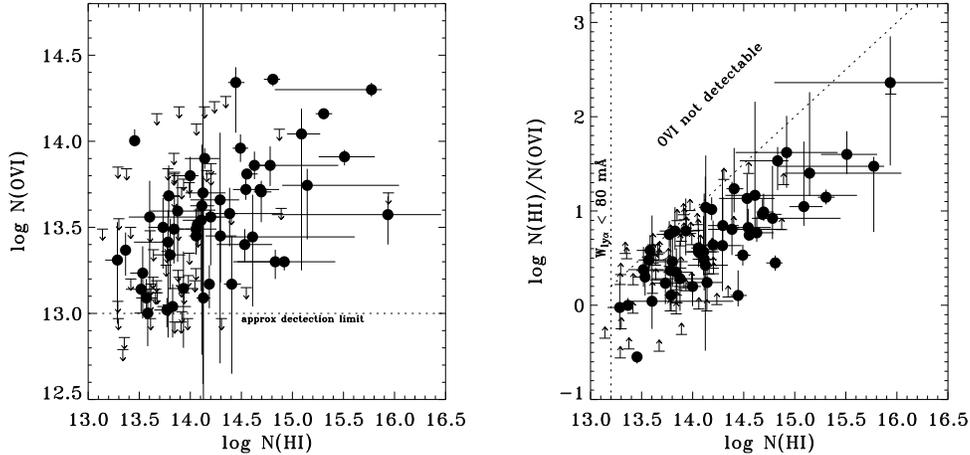}
\caption{We find mild correlation, if any, between H\,I and O\,VI column densities.  While N(H\,I) varies over a factor of nearly 1000, N(O\,VI) varies by a factor of only 30.  The ``multiphase ratio'' N(H\,I)/N(O\,VI) shows a dispersion of one dex, leading to an interpretation of the IGM with a neutral phase and a collisionally and photoionized hot phase.}
\end{figure}

\section{The Multiphase IGM}

O\,VI is an ideal tracer (at $10^{5.5}$ K) for the warm-hot ionized medium (WHIM; $10^5-10^7$ K) while H\,I composes the bulk of the warm neutral medium (WNM; $10^{3.5}-10^{4.5}$ K).  N(O\,VI) shows a mild correlation with N(H\,I) in the left panel of Figure~1.  While N(H\,I) varies by nearly 3 orders of magnitude, N(O\,VI) is detected between $10^{13}$ to a few times $10^{14}$ cm$^{-2}$ (a factor of $\sim$30).  The {\it ``multiphase ratio''} N(H\,I)/N(O\,VI) shown in the right panel of Figure~1 exhibits a moderate correlation with a dispersion of roughly one dex.

The large range in N(H\,I) and correlation in multiphase ratio imply that the IGM has at least two phases (WHIM and WNM).  We suggest that the warmer phase occupies a shell around a colder, neutral core of arbitrary size.  The outer WHIM shell is heated by a combination of external ionizing photons (from QSOs) and shocks from infalling clouds.  The narrow range of N(O\,VI) compared to N(H\,I) implies that the WHIM shell may have a characteristic size, while the neutral core can be arbitrarily large.

\begin{table}
\begin{center}
\caption{IGM Absorber Statistics}
\smallskip\smallskip
{\small
\begin{tabular}{ccccc}
\tableline
Absorbers & Criteria & $\cal{N}$ & $\Delta z$ & d$\cal{N}$/d$z$ \\
\tableline
H\,I & W$>$80 m\AA, $z<$0.15  & 126 & 2.08 & 61   \\
H\,I & W$>$240 m\AA, $z<$0.15 &  50 & 2.08 & 24   \\
\tableline
O\,VI & W$_{1032}\ga$12.5 m\AA\      &  52 & $<$2.0  & $>$25\\
O\,VI & W$_{1032}\ga$30 m\AA\        &  37 & 2.08 & 18   \\
O\,VI & W$_{1032}\ga$50 m\AA\        &  22 & 2.08 & 11   \\
\tableline\end{tabular}}\end{center}
\end{table}

\begin{figure}[t]
\plotone{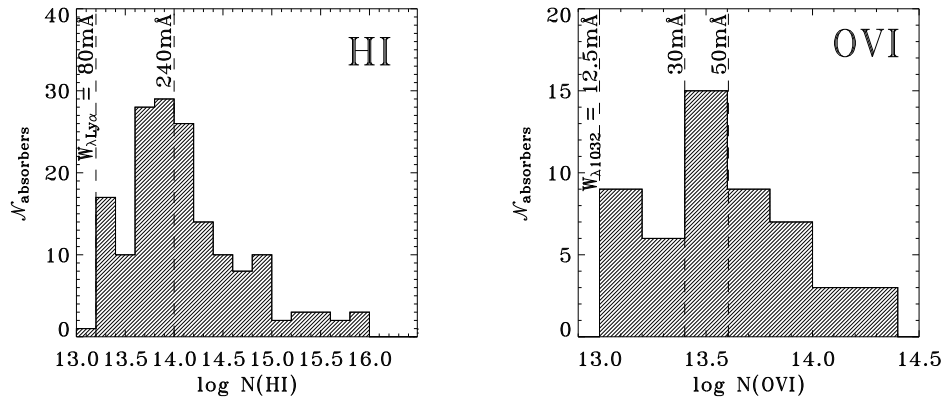}
\caption{Histogram of H\,I and O\,VI column densities in IGM absorbers.  Approximate W$_\lambda$ values for Ly$\alpha$ and O\,VI $\lambda$1032 are shown as vertical lines.}
\end{figure}

\section{The Hot Baryon Content of the Universe}

Using O\,VI as a tracer of the $10^{5-6}$ K portion of the WHIM, we can employ the number of absorbers per unit redshift to determine $\rm\Omega_{WHIM}$, the fraction of the universe's critical density made up by WHIM gas.  Our detection statistics are shown in Table~1 and Figure~2 for H\,I and O\,VI.  We estimate that our O\,VI survey is complete down to 30 m\AA\ in the 1032 \AA\ line (log\,N(O\,VI)$\sim$13.4) yielding 37 absorbers.  The total path length surveyed in the 26 sight lines is $\Delta z=2.08$ for a d$\cal{N}$/d$z=18\pm3$.  We assume H$\rm_o$=70 km s$^{-1}$ Mpc$^{-1}$, an O to H abundance of 10\% of the solar value ($4.9\times10^{-4}$; Allende Prieto et al. 2001), and an ionic fraction (O\,VI/O)=0.2 \citep{SutherlandDopita93}.  Using the equations in \citet{Savage02}, we get $\rm\Omega_{WHIM}=0.0022\pm0.0004$.  Uncertainties are based on Poisson statistics of $37\pm6$ absorbers.  It should be noted that this is a lower limit and the value is likely to go up as we refine our redshift pathlength values and take into account the incomplete sample of 15 absorbers with log\,N(O\,VI)$<$13.4.

Our result is entirely consistent with the result published by \citet{Savage02} of $\rm\Omega_{WHIM}=0.002$ based on only six O\,VI absorbers; our survey has six times more absorbers.  A similar study by \citet{Tripp04} based on 44 higher-redshift O\,VI absorbers ($z>0.12$) finds $\rm\Omega_{WHIM}=0.0018$.  Our value gives $\rm\Omega_{WHIM}/\Omega_{baryon}=0.048\pm0.010$ and shows that WHIM gas in the $10^{5-6}$ K range makes up at least 5\% of the baryonic mass in the local universe.

\end{document}